\documentclass{ws-p8-50x6-00}

\begin{document}

\title{\vspace*{-2cm} \mbox{} \hfill BI-TP 99/13\\
\mbox{} \hfill May 1999\\
\vspace*{1cm}
Lattice QCD at non-zero baryon number}

\author{O. Kaczmarek with J. Engels, F. Karsch, E. Laermann}

\address{Fakult\"at f\"ur Physik, Universit\"at Bielefeld, D-33615 Bielefeld,
  Germany\\
E-mail: okacz@physik.uni-bielefeld.de}


\maketitle

\abstracts{
We discuss the quenched limit of lattice QCD at
non-zero baryon number density. We find evidence for a mixed phase that becomes
broader with increasing baryon number. Although the action is explicitly
$Z(3)$ symmetric the Polyakov loop expectation value becomes non-zero already
in the low temperature phase. It indicates that the heavy quark potential stays
finite at large distances, i.e. the string between static quarks breaks at
non-zero baryon number density already in the hadronic phase. This behaviour is
validated by calculating the heavy quark potential using Polyakov loop
correlations.
}
\section{Introduction}
A quantitative analysis of QCD at non-zero density is important for our
understanding of the behaviour of dense matter as it is created in heavy ion
collisions and exists in the cosmological context.
While the QCD phase diagram is well known for vanishing
baryon density from lattice QCD, for the region of non-zero density only
qualitative features can be understood in terms of models (bags, percolation,
strings, ...) and approximations (resonance gas, perturbation theory,
instanton liquid, ...). Non-zero baryon number is usually introduced by a
non-zero chemical potential $\mu$\cite{Hasenfratz}$^,$\cite{Kogut}.
This leads to a break down of the
probabilistic interpretation of the path integral representation of the QCD
partition function as the fermion determinant gets complex\cite{Barbour}.\\
The static limit of QCD at non-zero chemical potential $\mu$ has been
formulated by Bender et al.\cite{Stam} and Blum et al.\cite{Blum}. It seems that in this case the first
order deconfinement transition of the $SU(3)$ gauge theory turns into a
crossover for arbitrarily small, non-zero values of the chemical
potential\cite{Blum}. Rather than introducing a non-vanishing chemical
potential, i.e. formulate QCD at non-vanishing baryon number density in the
grand canonical ensemble, one may go over to a canonical formulation of the
thermodynamics and fix directly the baryon number\cite{Miller}. This is
achieved by introducing an imaginary chemical
potential\cite{Miller}$^,$\cite{Roberge}
in the grand canonical partition function and
performing a Fourier integration to project onto the canonical partition function
for a given sector of fixed baryon number\cite{Miller}
\begin{eqnarray}
Z(B,T,V) = \frac{1}{2\pi} \int_0^{2\pi} \mathrm{ d}\phi e^{-iB\phi} Z(i\phi,T,V).
\end{eqnarray}
This leads to a well
defined quenched theory\cite{kacz} at fixed baryon number, which describes the
thermodynamics of gluons in the background of a non-zero number of static quark
sources, suitably arranged to obey Fermi statistics.
\section{Simulation of quenched QCD with non-zero baryon number}
\begin{figure}[t]
\begin{center}
\epsfxsize=20pc 
\epsfbox{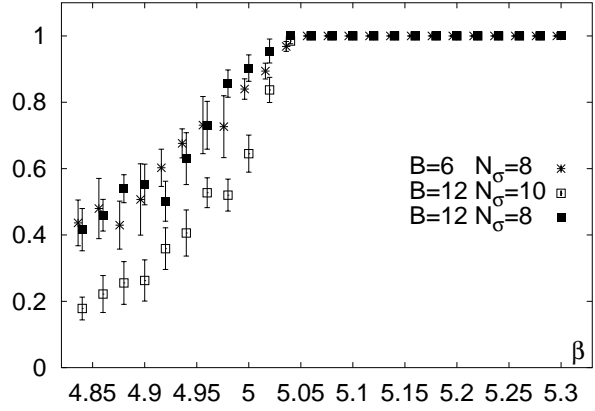} 
\caption{$\langle \mathrm{sgn}(\mathrm{ Re} \hat f_B)\rangle_{\vert\vert}$ for
  $B=6$ and 12 and lattices of size $8^3\times 2$ and $10^3\times 2$.}
\end{center}
\vspace*{-0.1cm}
\end{figure}
For any fixed value of the baryon number we can write the quenched partition
function as
\begin{eqnarray}
Z(B,T,V) = \int \prod_{x,\nu} \mathrm{ d}U_{x,\nu} \hat f_B e^{-S_G}
\end{eqnarray}
where the constraint on the baryon number is encoded in the function $\hat
f_B$ which is a function of Polyakov loops and $B$ counts the number of quarks,
i.e. $B/3$ is the baryon number. For $B=3$ $\hat f_B$ is, for instance, given
by\\
\begin{eqnarray}
f_{B=3} &=& (2\kappa)^{3N_\tau} ( V^3 \frac{4}{3} [ L_{1,0}]^3 \nonumber\\
&& + V^2 (8[ L_{1,0}][ L_{2,0} ] - 4[ L_{1,0}
][ L_{0,1} ] )\nonumber\\
&& + V (12 + \frac{2}{3} [ L_{3,0} ] - 2 [ L_{1,1} ] ) )
\end{eqnarray}
\hspace*{3cm} with $[ L_{i,j} ] = V^{-1} \sum_{\vec x} ({\rm Tr}
L_{\vec x} )^i ({\rm Tr} L^2_{\vec x})^j$.
\newpage
For a more detailed description of $\hat f_B$ see Engels et al.\cite{kacz}. $S_G$ is the
gluonic action, which is $Z(3)$ symmetric. The partition function $Z(B,T,V)$,
is non-zero only if $B$ is a multiple of 3, because $\hat f_B$ is invariant
under $Z(3)$ transformations only if $B$ is a multiple of 3. In general it
changes by a factor $e^{2\pi i B/3}$ under a global $Z(3)$ transformation of
timelike link variables.\\
\begin{figure}[t]
\begin{center}
\epsfxsize=13.8pc 
\epsfbox{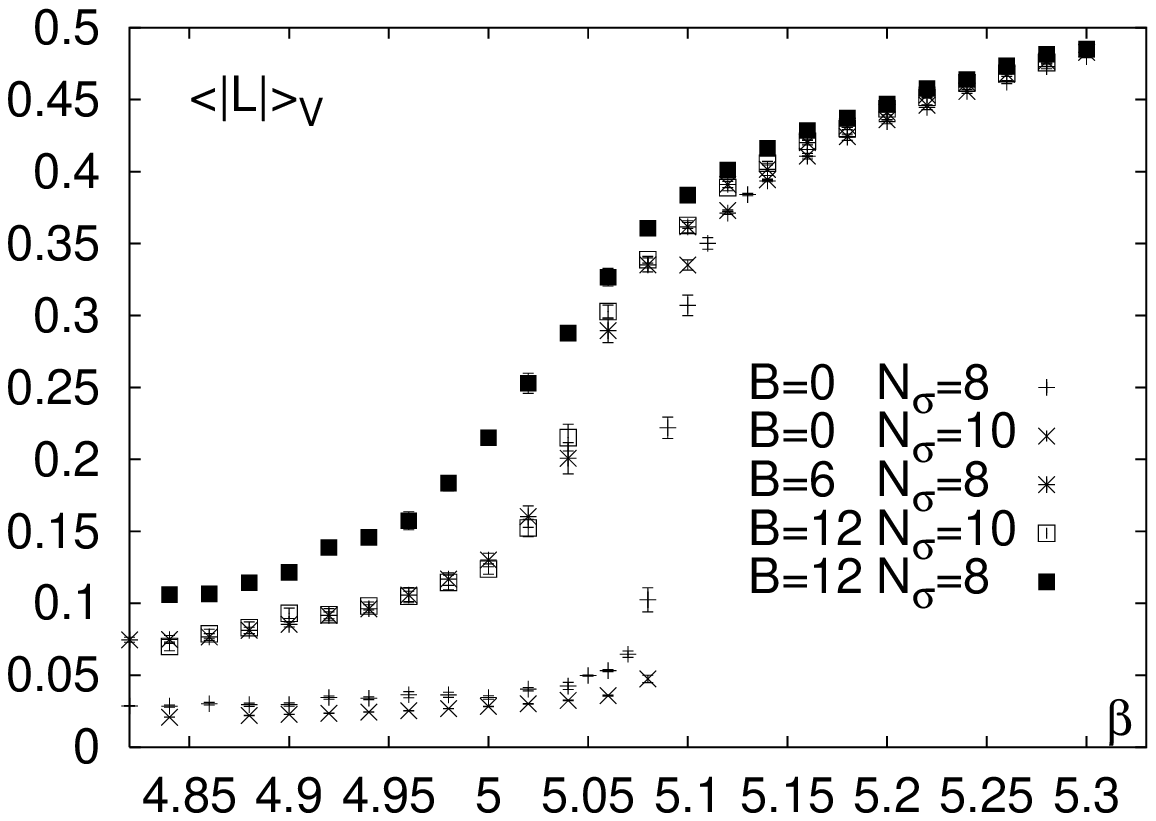} 
\epsfxsize=13.8pc 
\epsfbox{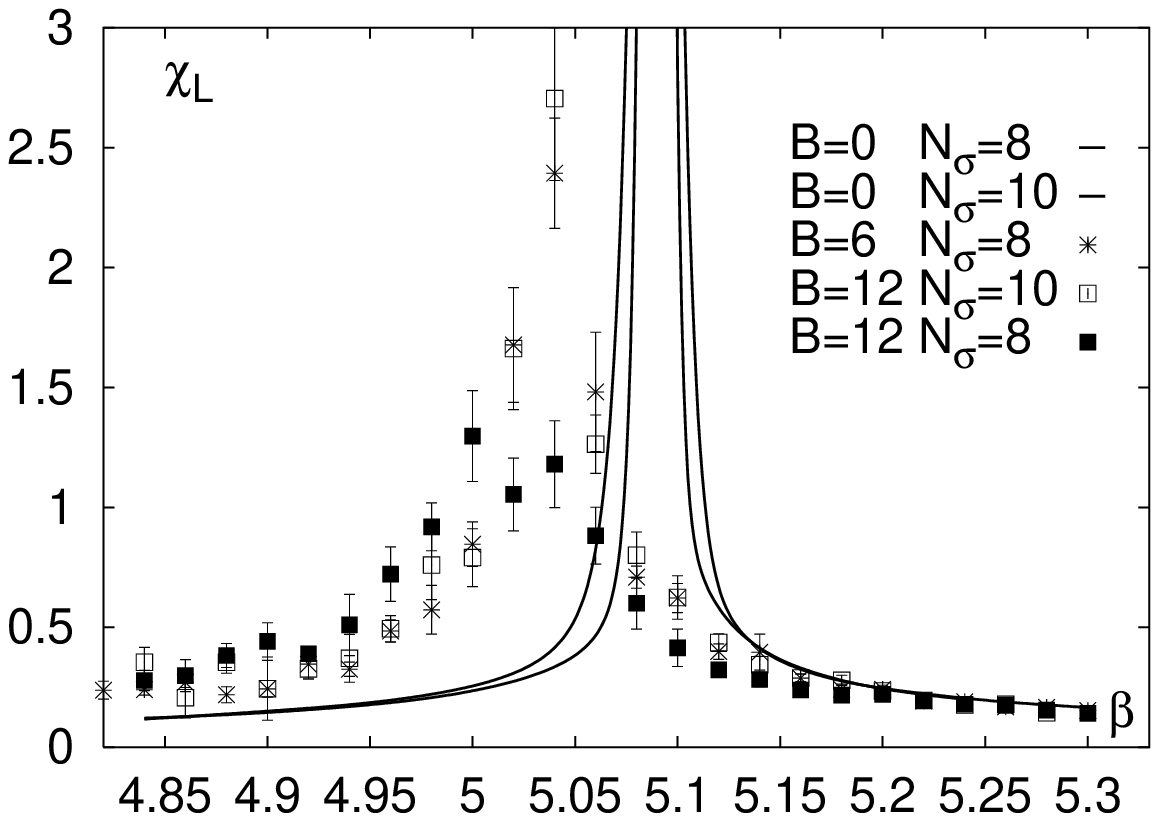} 
\caption{Polyakov loop expectation value (left) and Polyakov loop
  susceptibility (right) for different values of $B$ and
  lattices with spatial extend $N_\sigma=8$ and 10.}
\end{center}
\end{figure}
$\hat f_B$ is still a complex function, but upon integration over the gauge
fields the imaginary part of the partition function vanishes. The remaining sign problem can be
handled by including the sign in the calculation of observables\cite{Engels}.
We have performed simulations for the
one flavour case ($n_f=1$) using the partition function
\begin{eqnarray}
Z_{\vert\vert}(B,T,V) = \int \prod_{x,\nu} \mathrm{ d} U_{x,\nu} \vert
\mathrm{Re} \hat f_B \vert e^{-S_G}.
\end{eqnarray}
Expectation values of an observable ${\cal O}$ will be calculated according to
\begin{eqnarray}
\langle {\cal O} \rangle = \frac{\langle {\cal O} \cdot \mathrm{ sgn}
  (\mathrm{ Re} \hat f_B) \rangle_{\vert\vert}}{\langle \mathrm{ sgn}(\mathrm{
  Re}\hat f_B) \rangle_{\vert\vert}}.
\end{eqnarray}
The simulations are performed on $N_\sigma^3 \times N_\tau$ lattices
with $N_\sigma=8,10$ and $N_\tau=2$ using the
standard Wilson action with quark number values of $B=6$ and
12. When increasing the gauge coupling $\beta$ the temperature is increased
while $n_B/T^3=\frac{1}{3} B (\frac{N_\sigma}{N_\tau})^3$ is kept fixed.
Close to $T_c$ a simulation on a $8^3 \times 2$ lattice with $B=12$ corresponds to
$n_B\simeq 0.15/fm^3$, i.e. approximately nuclear matter density. Figure
1 shows the average sign $\langle \mathrm{ sgn}(\hat
f_B)\rangle_{\vert\vert}$ as a function of the coupling $\beta$.
For large values of the temperature the sign is almost always positive, but also for
the smallest temperature in our analysis the sign can be well determined. It
depends on the spatial volume $N_\sigma^3$ but varies little with $B$.

\begin{figure}[t]
\begin{center}
\epsfxsize=20pc 
\epsfbox{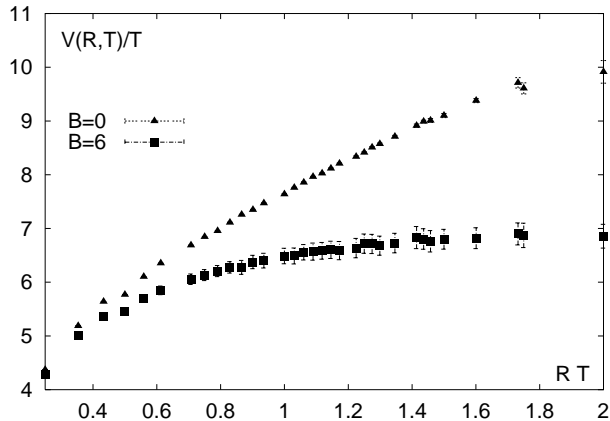} 
\caption{Heavy quark potential for $T\simeq 0.86 T_c$ and $B=0$ and 6.}
\end{center}
\end{figure}

In the quenched limit at vanishing baryon number the Polyakov loop expectation
$\langle L \rangle$
value is an order parameter for the deconfinement transition in the infinite
volume limit. For a first order phase transition, $\langle L \rangle$ changes
discontinuously at $T_c$.
In Figure 2 we see a clear signal for a first order transition for the $B=0$
case, while for all $B>0$ the transition is continuous. The transition
region is shifted towards
smaller values and it
broadens with increasing $B$. This is the expected behaviour for a
canonical calculation. By changing the gauge coupling $\beta$ we vary the
lattice cut-off and through this also the baryon number density
continuously. At fixed baryon number we therefore follow a simulation path that
traverses the mixed phase continuously.\\
If there is a mixed phase in the sense that there exist singularities in
thermodynamic observables when entering and leaving this mixed phase, it could
be reflected in a discontinuous change of the slope of the Polyakov loop
expectation value. We analyzed this by calculating the conventional Polyakov
loop susceptibility $\chi_L$ (Figure 2). This response function reflects the existence of
a transition region that becomes broader with increasing $n_B$, but does
not show indications for a discontinuity.

The Polyakov loop expectation value becomes non-zero already in the low
temperature phase. This indicates that the heavy quark potential stays finite
at large distances. We validate this by calculating the potential using
Polyakov loop correlations. The potentials for $T\simeq 0.86 T_c$ at
$B=0$ and 6 on a $16^3\times 4$ lattice are plotted in Figure 3. For zero
baryon number it shows the
usual behaviour for the quenched case. The potential is linearly rising for
large distances. For $B=6$ the potential stays finite at large distances.
The static quark anti-quark sources used to probe the heavy quark potential can
recombine with the already present static quarks. This leads to string
breaking even in the low temperature phase similarly to full QCD\cite{Edwin}.

The plaquette expectation value (Figure 4) shows a similar behaviour as the
Polyakov loop. With increasing $n_B$ the transition region broadens. Through an
integration over differences of plaquette expectation values for finite
temperature and zero temperature, the free energy
density can be calculated. As the area between these data increases with
increasing $n_B$, the free energy density decreases at fixed temperature with
increasing $B$.
\begin{figure}[t]
\begin{center}
\epsfxsize=13.8pc 
\epsfbox{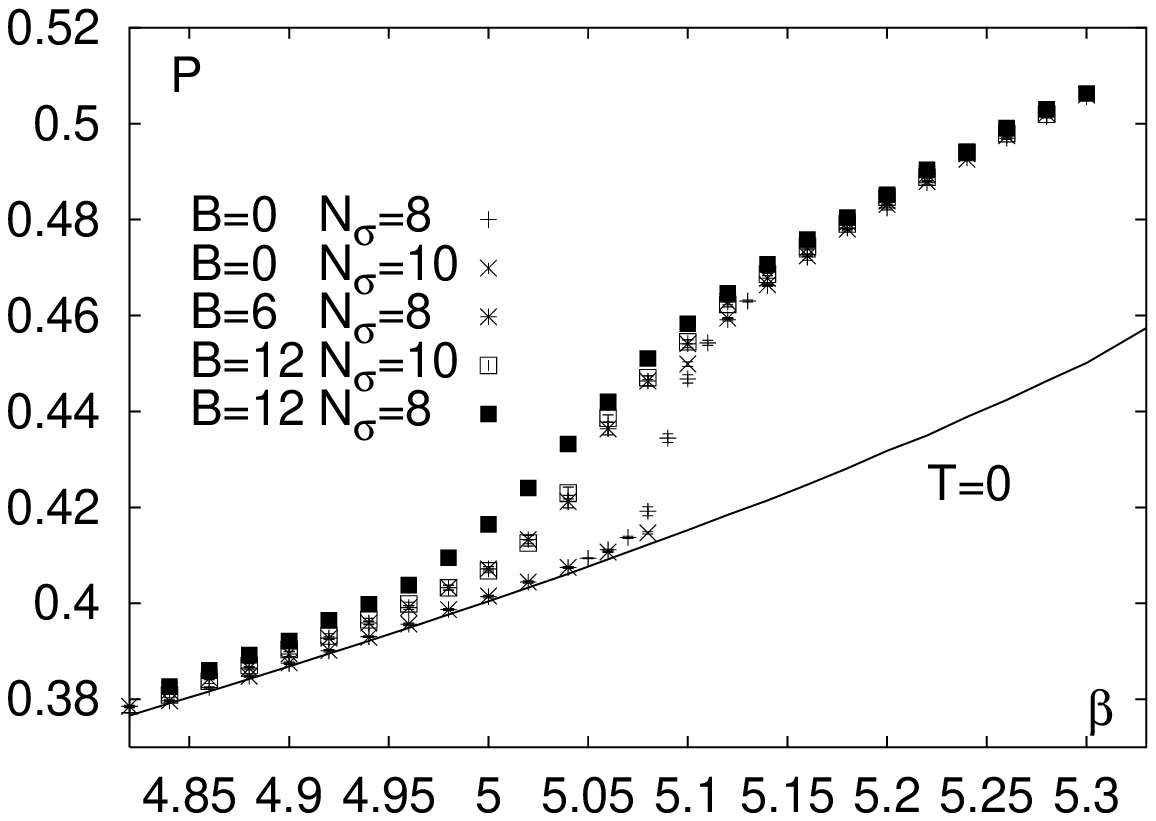} 
\epsfxsize=13.8pc 
\epsfbox{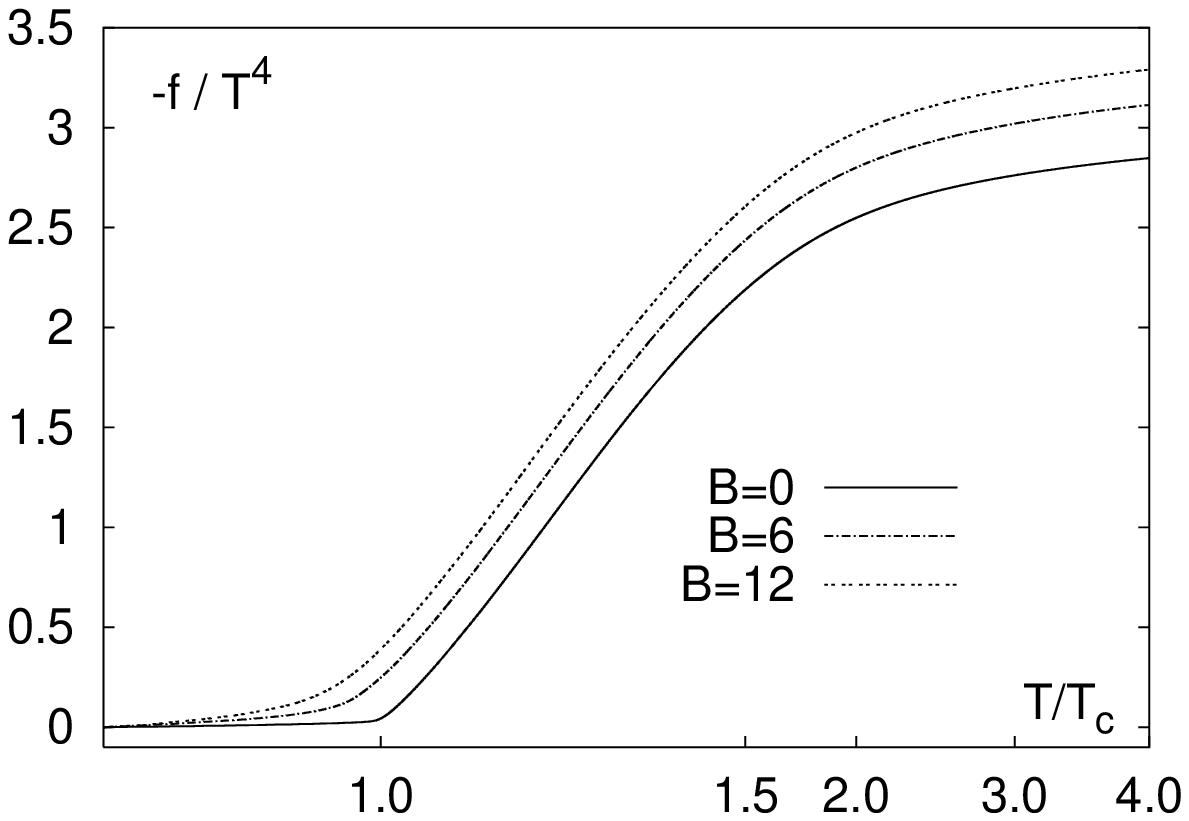} 
\caption{Plaquette expectation value (left) and the negative of the free energy
  density (right). On the right plot we only show tick marks at critical
  couplings on lattices with extend $N_\tau=2, 3, 4, 6$ and 8, which corresponds
  here to temperatures $T/T_c=1, 1.5, 2, 3, 4$}
\end{center}
\end{figure}

\section{Conclusions}
We have analyzed the quenched limit of QCD at non-zero baryon number. Although
a sign problem remains in this theory, it can be handled quite well
numerically. We find indications for a mixed phase, which broadens with
increasing baryon number density and is shifted towards smaller
temperatures. We also see evidence that the heavy quark potential stays finite
for large distances already in the hadronic phase. String breaking starts at
short distances.

\section*{Acknowledgments}
This work was partly supported by the TMR network {\it Finite Temperature Phase
  Transitions in Particle Physics}, EU contract no. ERBFMRX-CT97-0122 and the
  DFG through grant no. KA 1198/4-1.

\end{document}